# Fixational eye movements: about a binocular slow control mechanism

Marco Rusconi[a,b,#,*], Stephanie Jainta[c], Hazel Blythe[d], Ralf Engbert[b]

[a] Institute of Mathematics, DYCOS, University of Potsdam, Potsdam, Germany.
[b] Department of Psychology, University of Potsdam, Potsdam, Germany.
[#] Present address: Bernstein Center for Computational Neuroscience, Charité Universitätsmedizin Berlin, Berlin, Germany.
[c] IfADo - Leibniz-Institut für Arbeitsforschung, Dortmund, Germany.
[d] Department of Psychology, University of Southampton, Southampton, United Kingdom.

[*]Corresponding author:
Marco Rusconi
e-mail: dr.m.rusconi@gmail.com

**Abstract**
   Even when we look at stationary objects, involuntarily our eyes perform miniature movements and do not stand perfectly still. Such fixational eye movements (FEM) can be decomposed into at least two components: rapid microsaccades and slow (physiological) drift. Despite the general agreement that microsaccades have a central generating mechanism, the origin of drift is less clear. A direct approach to investigate whether drift is also centrally controlled or merely represents peripheral uncorrelated oculomotor noise is to quantify the statistical dependence between the velocity components of the FEM. Here we investigate the dependence between horizontal and vertical velocity components across the eyes during a visual fixation task with human observers. The results are compared with computer-generated surrogate time series containing only drift or only microsaccades. Our analyses show a binocular dependence between FEM velocity components predominantly due to drift. This result supports the existence of a central generating mechanism that modulates not only microsaccades but also drift and helps to explain the neuronal mechanism generating FEM.

**Introduction**

Even during fixating a stationary target our eyes perform erratic tiny movements. These fixational eye-movements (FEM) are the superposition of two types of motion[1]: microsaccades and drift. About 97% of the fixation time consists of drift movements (Yarbus, 1967), an eye-movement type whose underlying control mechanism (slow control) allows keeping the gaze relatively stationary on a fixed target (Steinman et al., 1973). Drift is interrupted by rare events, which are small-amplitude movements (< 1°) and almost always simultaneously occurring in both eyes (Krauskopf et al., 1960). Such microsaccades typically have a mean duration of about 25 ms (Ditchburn, 1980) and an average occurrence rate of 1-2 per second (Martinez-Conde et al., 2004).

Observed already in the 18th century (Jurin, 1738), the role of FEM is still not fully understood and their physiological function is still debated (Martinez-Conde et al. 2004, 2009; Rolfs, 2009a). Particularly, it is discussed whether the process that generates FEM might be central (e.g. located in the brain stem) or peripheral (noise in the motor neurons). While for microsaccades evidence converges on a central generator (Otero-Millan et al., 2008; Laubrock et al., 2008, 2010; Rolfs et al., 2006, 2008a, 2008b; Hafed et al., 2009a; Mergenthaler & Engbert, 2010; Engbert et al., 2011), drift was reported to be coherent (Spauschus et al., 1999) and conjugate (Ditchburn & Ginsborg, 1953) as well as independent (Yarbus, 1967) and uncorrelated (Krauskopf et al., 1960).

Recent experiments suggested the importance of drift for improving perception already at the early stage of visual processing (Rucci et al., 2007; Kuang et al., 2012). However, the origin of fixational drift remains unclear. On the one hand, drift might result from noisy firing of oculomotor neurons (Eizenman et al., 1985). On the other hand, drift might be centrally modulated based on a complex brain network including the superior colliculus (SC; see discussion in Rolfs, 2009b and Hafed et al. 2009b). Characterizing the statistical properties of drift movements (Engbert & Kliegl, 2004) can allow inferences on the neuronal origin of slow control and is thus important to understand the role of the ocular instability during fixation crucial for visual perception.

In order to understand the control principles underlying FEM, we systematically investigated the dependence across the eyes. In contrast to previous studies, we recorded FEM from a large sample of participants and computed the correlation of various FEM components across the eyes and two data sets recorded with two different eye-tracking systems (i.e., video-based and Dual-Purkinje eye-tracker). Based on experiments suggesting that the velocity of an image rather than its absolute position on the retina is the sensory input to the slow control of drift (Epelboim & Kowler, 1993), we used the eye velocities rather than the eye positions to analyze the dependence structure of the FEM components. Comparing the results with surrogate data, we aimed at understanding whether fixational drift movements are centrally modulated or merely uncorrelated motor noise.

**Materials and Methods**

Horizontal (H) and vertical (V) components of FEM were derived from binocular eye-movement trajectories recorded in a fixation experiment in which participants were instructed to fixate a target for several seconds without moving the eyes. To ensure that the obtained results did not depend on the video-based eye-tracking technique, we also analyzed binocular fixations recorded with two Dual-Purkinje Image (DPI) eye trackers.

*Video-based data.* The data were previously published with a different focus in Engbert and Mergenthaler (2006) and Mergenthaler and Engbert (2007). During a fixation experiment the horizontal and vertical components of FEM were recorded from both eyes while participants were sitting. In order to reduce movements, a chin-rest stabilized the head of the participants. Recording was performed in a dark room and luminance was kept constant. 20 students of the University of Potsdam with normal or corrected to normal vision participated in the experiment consisting of 30 trials. For some participants, a few trials had to be discarded resulting in four participants with 29 trials, one with 27 and one with 28 trials. Each trial consisted of 20 seconds of fixation on a static target (i.e., a dot on a uniform background) followed by 10 seconds of relaxation on natural landscape pictures. During recording trials interrupted by eye blinks were repeated. The eyes' trajectories were recorded with a high-speed video-based eye tracker (EyeLink-II, SR Research, Osgoode, ON, Canada) with 500 Hz sampling. The raw data were smoothed to reduce the presence of high frequency noise (Engbert & Mergenthaler, 2006), using a four-point velocity moving average (Engbert 2006).

*Dual-Purkinje data.* The full set of data was analyzed in depth and reported regarding a reading experiment aimed at investigating binocular impacts on word identification elsewhere (Jainta, Blythe, & Liversedge 2014). For the present analysis, we selected fixation data for N = 11 participants aged between 18–32 years with good visual acuity in each eye (better than 0.8 in decimal units). No participant wore glasses or contact lenses during the experiment. During recording with two DPI eye trackers (sampling rate 1000 Hz; spatial resolution <1 min arc) participants bit on a wax dental mold and used forehead rests to minimize head movements. In between sentence presentations, the task for the participants was to fixate a central dot (0.25 deg in diameter; presented at eye-height and at a viewing distance of 70 cm). Out of 48 trials per participant, a total number of 178 trials remained after excluding trials containing blinks or large saccades. Only the last saccade moving the eyes from the sentence-final word to the fixation dot remained in the data. Each analyzed epoch thus consisted of a saccade plus the following fixation leading to a total average duration of 1900 ms.

Finally, before computing the correlation coefficients, each epoch was band-pass filtered between 0.001 and 100 Hz (Eizenman, Hallett, & Frecker 1985) with a FIR filter of order n = 100 to remove noise above 100 Hz. The upper cut-off of 100 Hz was chosen based on FEM data recorded from an artificial eye in order to avoid spurious correlations across and within eye velocities that could only be due to high-frequency system noise. Filter design and application as well as the following analyses were done with standard Matlab (The MathWorks Inc., Natick, Massachusetts) tools. In order to avoid border effects and phase shift of the signals due to filtering, the initial n sample points (corresponding to the filter order) were removed.

*Data analysis.* The scope of the present analysis was to investigate the statistical dependence between horizontal and vertical components of FEM. As experimental evidence supports a velocity-based rather than a position-based analysis to characterize FEM (Epelboim & Kowler, 1993), we investigated the dependence between the velocity components of FEM rather than the position components. Velocities were computed as increments of the smoothed time series after removing the mean. The dependence was quantified by means of Spearman's $\rho$. The choice of using a rank-based instead of a standard correlation analysis was motivated by several properties of the FEM. Firstly, the time series of FEM contained microsaccades. Fig. 1 shows that microsaccades (red dots) are essentially velocity outliers with respect to the slow background motion activity, and it is well known that the Pearson correlation coefficient *r* is extremely responsive to outliers (Abdullah, 1990). As a matter of fact, the occurrence of binocular microsaccades contributed to a (positive) Pearson correlation, overestimating the actual dependence. In addition, the number of microsaccades and their amplitude differed in horizontal and vertical components. Thus, microsaccades could even have a differential effect on the correlations between velocity components. Secondly, drift velocities are not normally distributed (Cherici, Kuang, Poletti, & Rucci 2012) and Pearson's *r* is not robust against non-normality (Kowalski, 1972). Finally, the rank-based Spearman's $\rho$ is more general than the Pearson's *r* because it captures any monotonic dependence in the data and not only linear trends.

We analyzed the dependencies between the horizontal left and horizontal right component, and between the vertical left and vertical right component of the FEM. We refer to these components as parallel components. For each trial of each participant we computed the Spearman's $\rho$ between the velocity components. Averaging across trials, the corresponding mean value $\bar{\rho}$ characterizes the statistical dependence for each participant. The correlation coefficients are not normally distributed. For this reason, before averaging them or computing the relative confidence interval (ci), the correlation values were transformed using Fisher's z-transform (Fisher, 1915, 1921). After computing mean and ci, the corresponding values were transformed back into meaningful velocities. In the following, we will report the velocities only.

In a second analysis, we removed the microsaccades from the FEM time series to examine a possible effect only of drift on the dependence between the eyes. This has been done using a two-step objective procedure. In a first step, microsaccade onsets and offsets have been detected with a 2D velocity-threshold algorithm as in Engbert and Kliegl (2003; see also Engbert & Mergenthaler, 2006). After converting position-based trajectories into velocity time series, microsaccades were detected as velocity outliers (i.e., all events exceeding an elliptic threshold relative to noise level). Similar to Engbert and Mergenthaler (2006), the threshold parameter was set to $\lambda = 5$ multiples of the standard deviation of each velocity component. Finally, a minimum duration criterion was adopted to select only events with durations larger than 6 ms. Finally, the epochs between onset and offset of the detected microsaccades were removed from the velocity time series.

To examine the effect of only the microsaccades on the dependence between the velocity components, the correlation analysis was also performed on phase-randomized amplitude-adjusted surrogate data (Theiler et al., 1992; Engbert & Mergenthaler, 2006). To generate the surrogate data, the velocity time series without microsaccades were shuffled. After random shuffling the Fourier phases, the microsaccadic epochs were inserted at the original time positions. This technique ensured that (i) the velocity distribution of FEM remained unchanged and that (ii) the autocorrelation structure of the surrogate time series approximated the autocorrelation of the original data.

*Ethical standard*

Concerning the data collected with the DPI eye trackers, the experimental procedure was approved by the University of Southampton Ethics and Research Governance Office and followed the conventions of the Declaration of Helsinki. Informed written consent was obtained from each participant after explanation of the procedure of the experiment. The video-based data were already reported with a different focus in Engbert and Mergenthaler (2006) and Mergenthaler and Engbert (2007).

*Statistical Methods*

The data are reported as mean value $\pm$ confidence interval. The confidence interval is computed as $1.96 \frac{s}{\sqrt[2]{N}}$, where N is the corresponding number of values entering the mean. Before averaging the correlation coefficients or computing the relative confidence interval, we applied the Fisher's z-transform (Fisher, 1915, 1921) to correlation values. The plots show the inverse transformed average and corresponding confidence interval. To access the significance of the effect of removing microsaccades we used a paired t-test.

**Results**

To test the hypothesis that FEM are independent oculomotor noise, we computed Spearman's ρ for the parallel velocity components across both eyes. Fig. 2 (panels A and D) shows the mean statistical dependence for the horizontal and the vertical velocity components of the FEM (i.e., between the parallel components in the left and right eye). The entire time series was considered for this analysis including microsaccadic events. The bars correspond to the value of $\bar{\rho}$ obtained for each participant. The participants were ordered by microsaccade rates with the rates reported on the abscissa. The analysis showed that the parallel velocity components were dependent across the eyes. For all participants, in both the horizontal and the vertical components, the dependence was always positive, and participants with a higher rate of microsaccades had larger values of $\bar{\rho}$. This result shows the existence of a monotonically increasing mapping between the parallel velocity components of FEM and stands in contrast to the independence postulated under the oculomotor noise assumption.

Fig. 2 (panels B and E) shows the values of $\bar{\rho}$ obtained after removing the microsaccades. Contrary to the hypothesis of independent and uncorrelated processes, the velocity components of the remaining drift epochs were still correlated, and this rank-correlation was positive. Again, participants with a higher microsaccade rate showed the larger correlations, especially between the horizontal velocities. This analysis indicates dependence between the parallel velocity components in case of drift even if microsaccades are removed.

To further quantify the impact of removing the microsaccades on the measured dependence, we tested for differences between the mean $\langle\rho\rangle$ across participants of the rank coefficients with and without microsaccades. Significance was assessed with a paired t-test on the z-transformed data. After removing the microsaccades we observed a small significant reduction of $\langle\rho\rangle$ in both horizontal and vertical components (mean difference < 0.05 for both, $t_{19}$ = 6.9888, $P$ = 1.17·$10^{-06}$ and $t_{19}$ = 5.5507, $P$ = 2.36·$10^{-05}$ respectively for H and V). From this test we can conclude that removing microsaccades have a small impact on the rank-correlations.

Finally, Fig. 2 (panels C and F) shows the values of $\bar{\rho}$ obtained for the velocity components of surrogate data that were obtained by constrained random shuffling (see Materials and Methods). This technique ensured that the surrogate time series contained microsaccades at the same temporal position as the original data, while randomly shuffled epochs of drift were generated between microsaccades. For the large majority of the analyzed participants the surrogate time series were correlated. However, microsaccades could not have had a strong impact on the dependence obtained in the entire FEM time series (including drift and microsaccades) because the absolute mean rank-correlations $\bar{\rho}$ of the microsaccades alone were small and not larger than 0.15.

To ensure that these findings were not due to artifacts of data processing, we further tested whether the correlations measured in the data were possibly due to the algorithm used to detect microsaccades. Microsaccades were detected with the velocity-threshold algorithm developed by Engbert and Kliegl (2003; see also Engbert & Mergenthaler, 2006). For the analysis presented in Fig. 2 we used $\lambda$ = 5 for the threshold, which was suggested by Engbert and Mergenthaler (2006) to be optimal. Increasing the value of $\lambda$ corresponds to a more conservative criterion for selecting microsaccadic events, and only microsaccades would be considered which are extreme outliers. As a consequence, after removing those microsaccades, more binocular events would be retained in the dataset that could contribute to dependencies. Conversely, by decreasing the value of $\lambda$ a larger part of the drift motion would be detected as microsaccades and thus removed. Fig. 3 reports the correlations observed for different values of $\lambda$. Decreasing $\lambda$, we observed the expected reduction of the dependence due to loss of data points. However, even for the extreme value of $\lambda$ = 2 we still found the critical positive correlations.

A last possible source of confound might have been that microsaccades were only removed partially leaving the residual part in the velocity components. To test this, we removed a larger portion of data adding an extra $\Delta t$ before and after each microsaccade onset and offset. Fig. 3 shows an example of the correlations obtained for $\Delta t$ = 160 ms and $\lambda$ = 5. While for some participants we observed a reduction of correlations, the global trend remained unchanged. These analyses suggest that the observed correlations were not due to undetected microsaccades or residual microsaccadic epochs that may have still been contained in the drift time series.

Finally, to test the robustness of our results, we replicated the same analyses with a different data set of FEM recorded with two DPI eye-trackers. Fig. 4 summarizes the results of these additional analyses and shows the comparison of the grand mean ⟨ρ⟩ of video-based and DPI data. Overall, the DPI data confirmed the positive correlation between parallel FEM components (see Fig. 4 B). Again, detection of microsaccades was done with λ = 5. Using this value, we not only removed the microsaccades but also the large horizontal saccade from the sentence-final word to the central fixation dot at the beginning of each fixation trial. Additionally removing the data belonging to this large saccade, we expected to see a large impact on the dependence between the eyes velocity in the drift time series. Indeed, we observed a substantial decrease in the correlations of the horizontal component in the drift time series after removing the (micro-) saccadic events. The impact of the large horizontal saccade could also be seen in the results of the surrogate data (MS time series in Fig. 4 B): the horizontal component showed a larger correlation than the vertical one. Despite these differences, the analysis of DPI data confirmed the existence of positive correlations due to drift. Finally, averaging across the horizontal and vertical correlations indicated that the video-based and DPI data showed a similar trend (see Fig. 4 C). Therefore, the DPI data supports the results obtained for eye movements recorded with the video-based system suggesting that FEM velocities are rank-correlated across the eyes and that such dependence is mostly due to drift rather than to microsaccades.

**Discussion**

The fundamental question we addressed in this study was whether the eyes' drift movements during fixation represent uncorrelated oculomotor noise. If, on the contrary, drift movements were correlated across the eyes this could indicate a central modulation of slow control guiding the drift. To study the coordination of FEM across the eyes we examined the correlation between the parallel velocity components that is between the horizontal component of the left eye and the horizontal component of the right eye, and likewise between the vertical components across the two eyes. All coefficients showed a positive correlation of the fixational velocity suggesting that FEM across eyes are not statistically independent but are yoked eye movements. In order to test whether drift and microsaccades contributed differently to the dependence, we recomputed the correlations after removing microsaccades from the time series, and for surrogate data sets containing only the microsaccades with randomly shuffled drift epochs. Most importantly, even after removing microsaccades from the time series we still observed a positive dependence between parallel components of the eyes. This indicates that the correlation between FEM velocities is mainly due to drift, and microsaccades contribute to the binocular dependence to a much smaller degree.

In addition, we obtained that higher correlations between the velocity components of FEM corresponded to a higher rate of microsaccades, independent of whether microsaccades were still present in the data sets or not. Participants with a higher microsaccade rate showed FEM that were more strongly rank-correlated. This result supports the idea that microsaccades are not only episodes embedded in some sort of background activity due to slow control, but that microsaccades and drift are dynamically related (Engbert & Kliegl, 2004; Engbert & Mergenthaler, 2006). As such, the present results are in good agreement with a model introduced by Engbert et al. (2011) that unifies the generation of both microsaccades and drift under a common mechanism.

The present evidence for binocular dependence contrasts with earlier findings of uncorrelated drift (Ditchburn & Ginsborg, 1953; Krauskopf et al., 1960; Yarbus, 1967). One possible explanation of this inconsistency is that the results reported previously were based on the analysis of only two participants (Ditchburn & Ginsborg, 1953; Krauskopf et al., 1960). Our much larger sample ($N_{tot}$ = 20 + 11) showed a large interindividual variability and, contrary to the previous studies, permitted to derive a more robust estimate of the population correlation.

More importantly, the finding of binocular drift dependence is also incompatible with the hypothesis of independent drift that is in principle expected from an oculomotor noise perspective. Nevertheless, King and Zhou (2000; see also Zhou & King, 1998) recently introduced a physiological model for eye movement control in which monocular pre-motor commands are modulated by binocular motoneurons resulting in the binocular coordination of the eyes. In this model, drift movements could be accounted for by correlated noise generated by the random firing of the binocular motoneurons. Given such correlated motor noise, a correlation between parallel FEM components would not necessarily rule out an oculomotor noise explanation in favor of a common central mechanism. However, Di Stasi et al. (2013) reported results that speak against the interpretation of drift as peripheral noise. They showed that drift is modulated during cognitive tasks due to mental fatigue. Thus, evidence is converging in favor of drift being centrally modulated rather than being pure motor noise.

But where could such a common central mechanism be located? Due to the importance of the SC in smooth pursuit eye movements, Rolfs (2009b) in a comment to Hafed et al. (2009a) suggested that the SC might play a role not only in the control of microsaccades but also in the modulation of drift. In response, Hafed et al. (2009b) argued that the modulation of SC activity during smooth pursuit could be due to the motion of the pursued target projected on the retina rather than the signature of pre-motor modulation of drift. Nevertheless, Hafed and colleagues agreed that drift seems to be controlled by a brain network that may possibly include the SC. This interpretation was supported by Di Stasi et al. (2013) who suggested an involvement of the brainstem in the modulation of the kinematic properties of FEM (drift and microsaccades). Although the neuronal mechanisms behind the control of drift may still not be fully identified, our results indicate that fixational drift is not just uncorrelated oculomotor noise. The evidence for binocularly correlated drift is in line with accounts that favor a central modulation of drift. The SC seems to be one of the best candidates among the neuronal circuitries to be involved in the slow control of fixational drift movements.

Recently, video-based recording systems were criticized because of the noise level during recording (Collewijn & Kowler, 2008) and because they may be prone to possible artifacts due to pupil oscillations during recording (Wyatt, 2010; Kimmel, Mammo, & Newsome, 2012). Concerning the effect of noise we would like to distinguish between the intrinsic noise of the recording system as discussed by Collewijn and Kowler (2008) and possible artificial signals introduced during recording or analysis due to external sources. As it has been done in the present study, intrinsic eye-tracking noise can be significantly reduced during analysis using signal-processing techniques such as smoothing. Note that in the present case, however, the applied smoothing cannot have generated spurious results because, although smoothing introduces an autocorrelation within the time series, it cannot generate artificial correlations between data sets given they are uncorrelated. Thus, smoothing cannot, even partially, explain our findings. Moreover, we presented a correlation analysis between signals (i.e., the velocity components) that are recorded by (in principle) independent sensors (i.e., the left and right camera). Thus, possible recording noise should be uncorrelated and have a small, if not irrelevant, effect on the observed correlations.

Different is the case of systematic signals artificially introduced during recording. Noise per-se can have a positive effect not only on physical systems (Gammaitoni, Hänggi, Jung, & Marchesoni, 1998), but also on biological systems (Wiesenfeld, & Moss, 1995; Hänggi, 2002; Moss, Ward, & Sannita, 2004; Rusconi, Zaikin, Marwan, & Kurths, 2008; McDonnell & Abbott, 2009) and on visual perception (Simonotto, Riani, & Seife, 1997). However, systematic noise as due to pupil oscillations or artifacts produced by head or eye-tracker movements may have the effect of masking the genuine FEM and may contaminate correlations. To reduce pupil confounds, we controlled for luminance conditions. No luminance variations were possible during recording and thus, the luminance-induced pupil oscillations described above can be excluded. A chin rest and head-movement correction was used, and a trial-by-trial calibration limited the presence of slow signals due to head movements or slip of the eye-tracker. Although the data were carefully recorded, it is not possible to exclude completely that the obtained correlations were biased by artificial signals such as other pupil oscillations for which we could not control. To clear any doubt on this matter, we analyzed binocular fixations recorded with a DPI eye-tracking system that allows high resolution recordings (Rucci, Iovin, Poletti, & Santini, 2007; Poletti, M., Listorti, C., & Rucci, 2010; Kuang, Poletti, Victor, & Rucci, 2012; Cherici, Kuang, Poletti, & Rucci 2012), that is typically used to study drift and that cannot be biased by pupil oscillations. Also in the DPI data, we observed the positive correlations between both horizontal and vertical velocity components and thus replicated the findings of binocular drift dependence.

Experimental and theoretical evidence seems to converge towards the existence of a central mechanism generating and/or partially modulating both motion types of FEM, drift and microsaccades. Our result is in line with this interpretation. However, in the case of drift, the alternative explanation involving binocular motoneurons is not completely ruled out. Microsaccades seem to be regulated by higher-level cognitive processes such as attention (Hafed & Clark, 2002; Engbert & Kliegl, 2003; Laubrock et al., 2005; Engbert, 2006; Laubrock et al., 2007, 2010). Therefore, a way to further confirm that slow control is centrally modulated may be to study whether drift is also affected by cognition. We think that future investigation in this direction is important to clarify the mechanisms generating and modulating FEM. Such research will be crucial for a better understanding of visual perception and visual stability.

**Acknowledgment**

The authors would like to thank Simon Liversedge for providing the DPI data; we would also like to thank Carsten Allefeld, Matthias Holschneider, and Konstantin Mergenthaler for helpful discussions and suggestions. Special thanks to Sarah Risse for her comments on an earlier version of the manuscript. MR received support from Deutsche Forschungsgemeinschaft: Research Unit 868, grant EN471/3 and project KFO 247.

Footnote

[1] A third type, tremor, is an aperiodic oscillation of the eyes that, due to its very small amplitude (Martinez-Conde et al., 2004), cannot be investigated with the techniques used here.

**Figure legends**

**Figure 1:** Scatterplot of velocities within the same eye (left panel) and across the two eyes (right panel). Data is presented for one example trial of one participant. In both panels we observe the presence of outliers due to microsaccades highlighted in red. The binocularity of the microsaccades introduces linear correlations between the parallel velocities, i.e., across eyes.

**Figure 2:** Dependence between the velocity components of FEM. Individual bars for each participant. Participants are ordered by microsaccade rate. Black error bars represent the 95% confidence interval of the mean of each participants (i.e. $1.96 \frac{s}{\sqrt[2]{N}}$, with s for the inverse Fisher-transform of the sample standard deviation of the Fisher-transformed means and N the number of trials for each participant). Panels A and D show the mean Spearman's r $\bar{\rho}$ for the original data for horizontal (FEM H) and vertical (FEM V) components respectively. For all participants we obtained a positive correlation $\bar{\rho}$, thus the velocity components of FEM are not independent between the eyes. The same is valid after removing microsaccades (Drift H and Drift V in panels B and E). Microsaccades alone (MS H and MS V in panels C and F) produced overall small correlations: the largest mean $\bar{\rho}$ was smaller than 0.15.

**Figure 3**: Effect of λ on the mean Spearman's $\bar{\rho}$ between horizontal (A) and vertical velocities (B). Green and red symbols show $\bar{\rho}$ obtained for λ = 2 and λ = 8. The blue bars correspond to λ = 5. Reducing λ lead to a reduction of dependencies, but even for λ = 2 we still observed positive $\bar{\rho}$ for all participants.

**Figure 4**: Group level analysis and contrast between the results with video-based (N = 20) and DPI (N = 11) data. Panel A shows the grand mean $\langle \rho \rangle$ for the horizontal (H) and vertical (V) velocity components. The correlations are compared between three data sets: (1) containing the full fixational eye-movement trajectory (FEM), (2) containing mainly the drift epochs with the microsaccades being removed (Drift), and (3) containing only the microsaccades with the drift epochs being randomly shuffled (MS). The data were recorded with a video-based system. Panel B shows the same for DPI data. In both cases we obtained a positive significant correlation. The average between horizontal and vertical components showed a similar trend for video-based and DPI data (panel C).

Figure 1

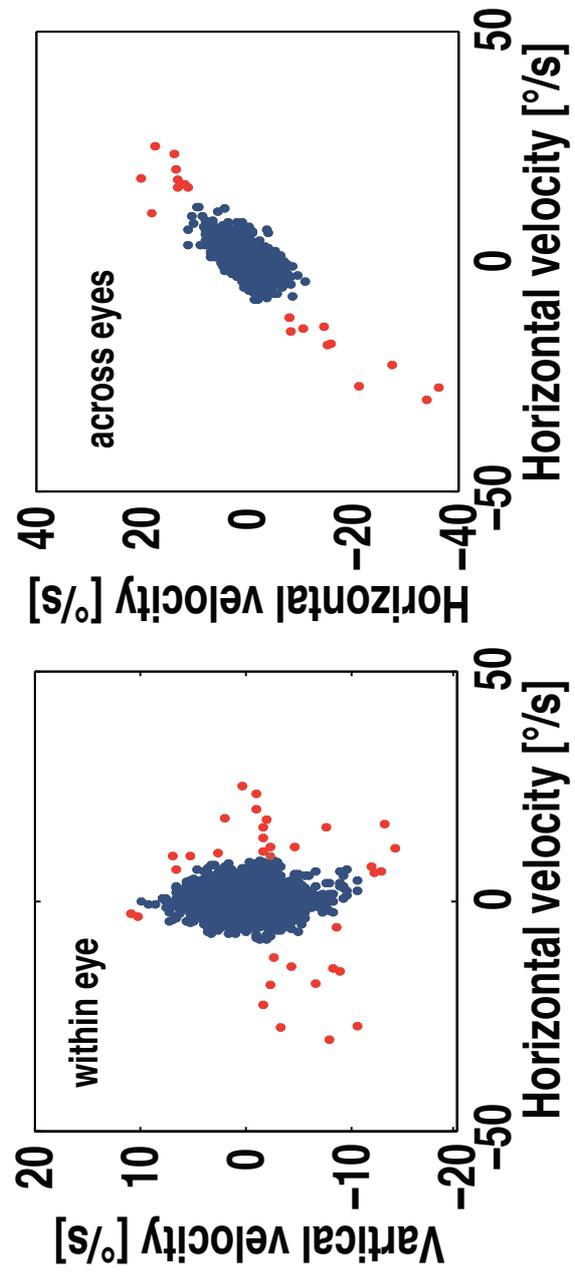

Figure 2

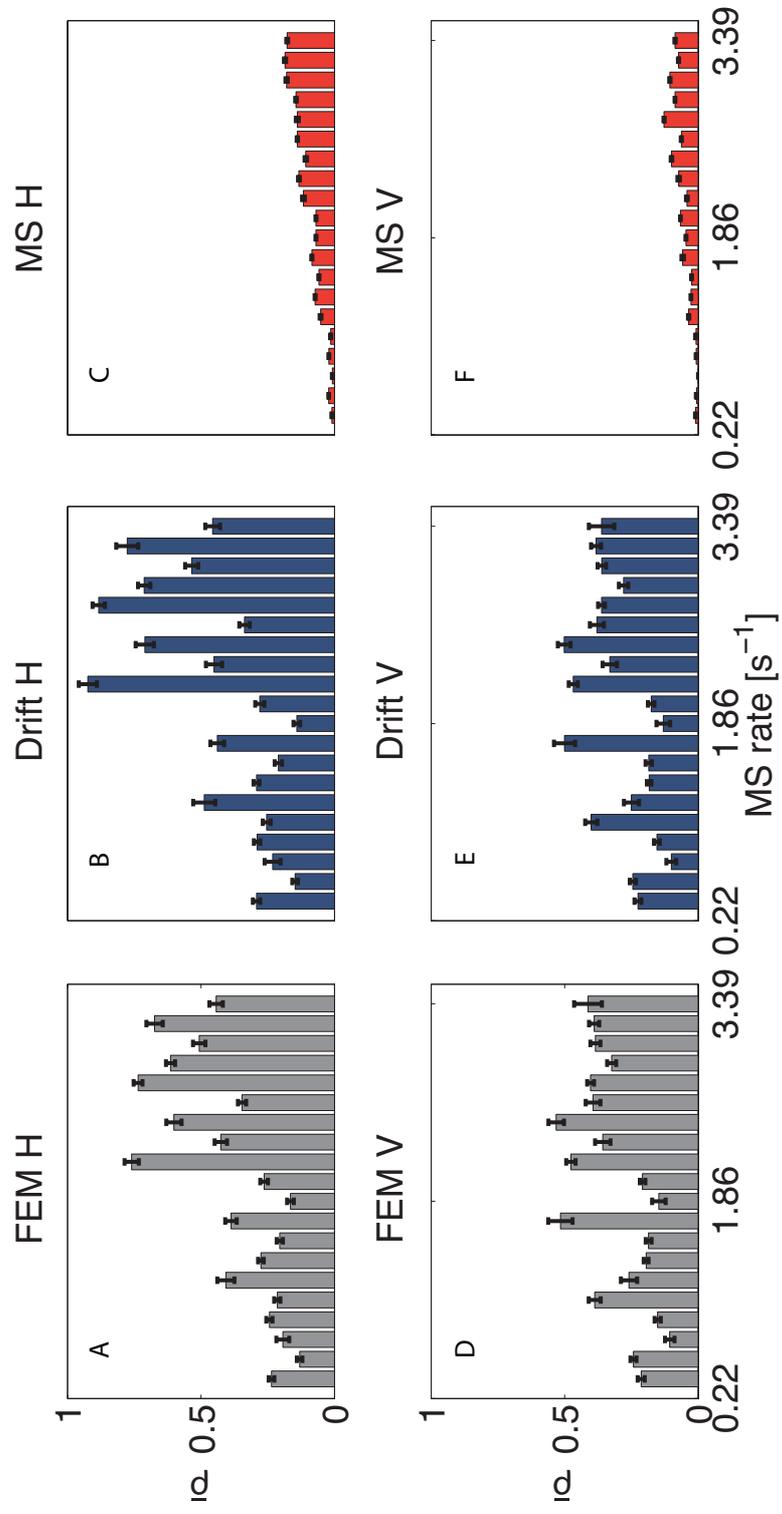

Figure 3

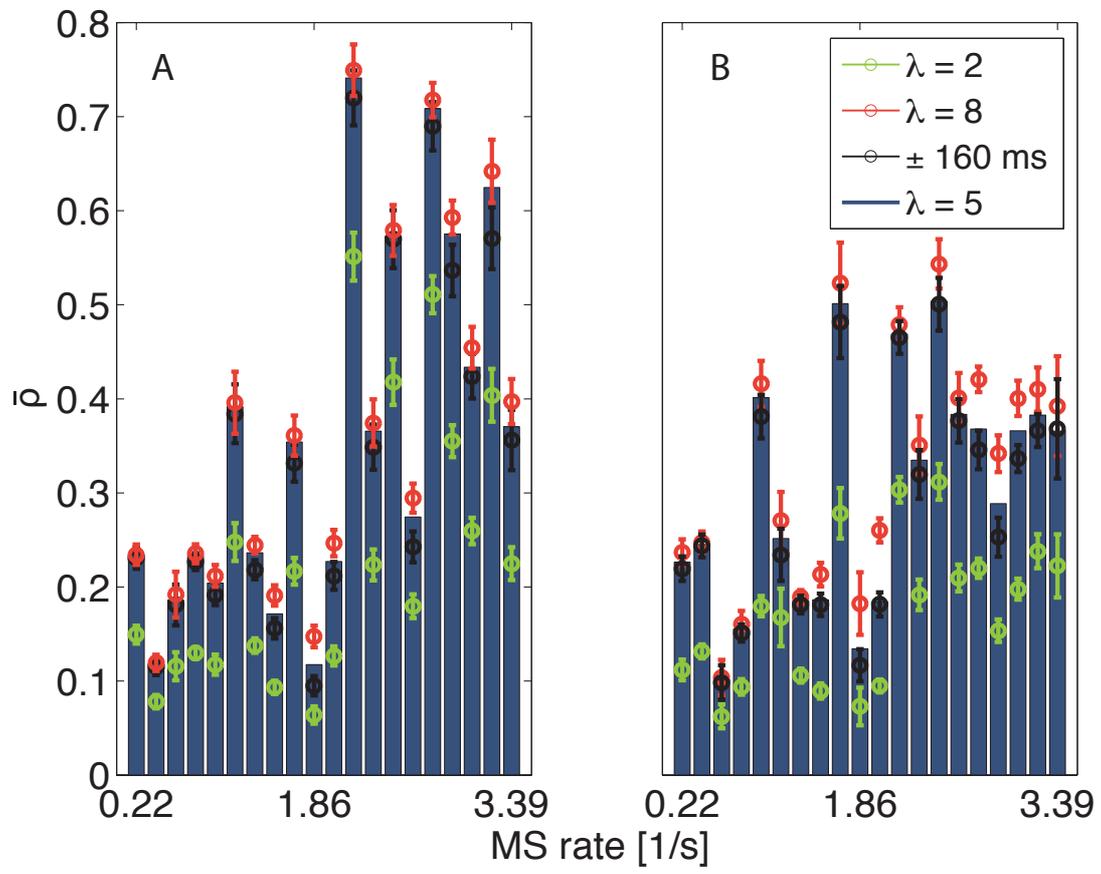

Figure 4

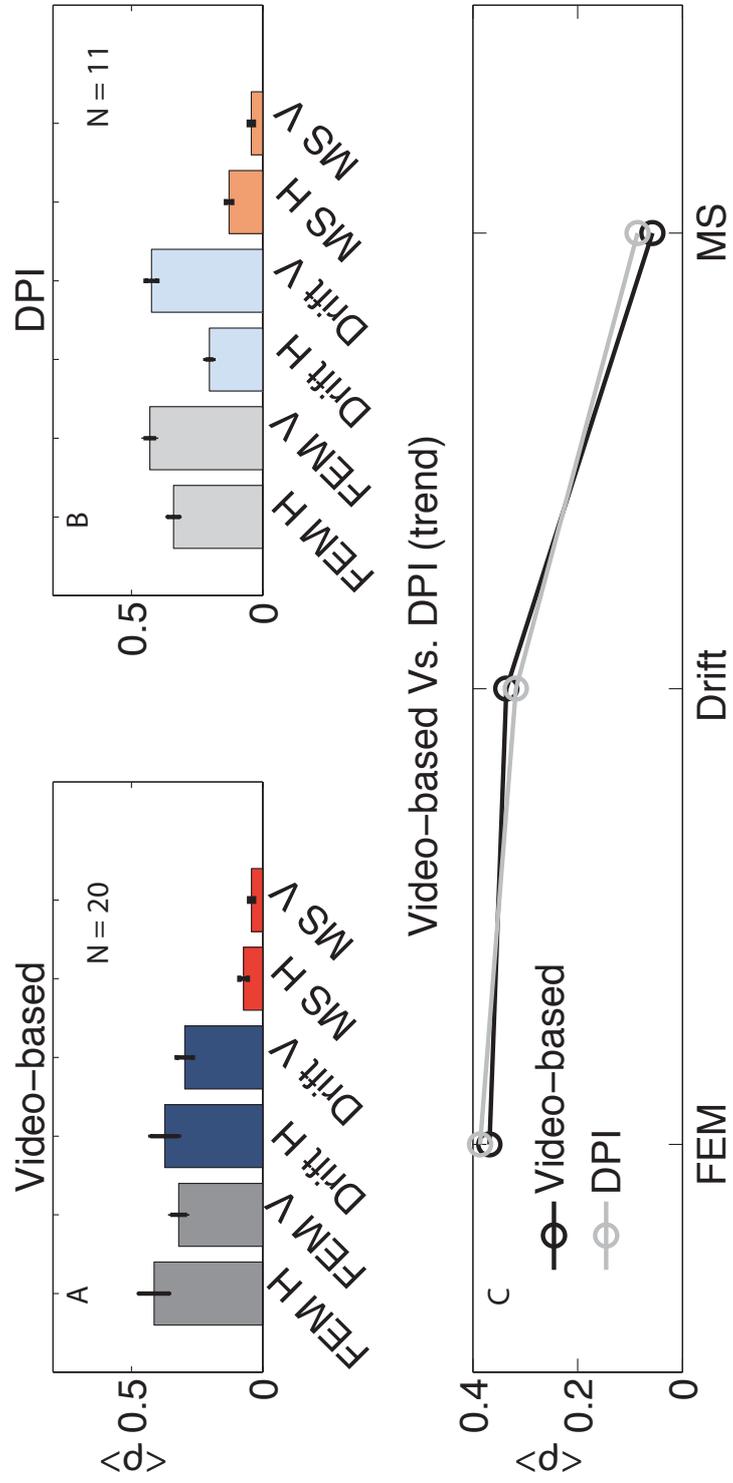